\documentclass[aps,pra,epsfigure,twocolumn,longbibliography]{revtex4-1}
\usepackage{dcolumn}    
\usepackage{bm} 
\usepackage{graphicx}
\usepackage{amsmath}    
\usepackage{latexsym}
\usepackage{amsfonts}   
\usepackage{amssymb}
\usepackage{array}      
\usepackage{epsfig}
\usepackage{txfonts}
\usepackage{color}
\usepackage[colorlinks=true,linkcolor=blue,urlcolor=blue,citecolor=blue,pdfusetitle]{hyperref}
\usepackage{hyperref}

\newcommand{\ket}[1]{\left\vert#1\right\rangle}
\newcommand{\bra}[1]{\left\langle#1\right\vert}

\newcommand{\blah}{blah\\blah\\blah\\blah\\blah.}

\catcode`\|=\active \def|{
\fontencoding{T1}\selectfont\symbol{124}\fontencoding{\encodingdefault}}

\begin{document}
 \title{Robust multipartite entanglement generation via a collision model}
       
\author{Bar\i\c{s} \c{C}akmak,$^{1,2}$ Steve Campbell,$^{3}$ Bassano Vacchini,$^{4,3}$, \"{O}zg\"{u}r E. M\"{u}stecapl{\i}o\u{g}lu,$^1$ and Mauro Paternostro$^5$}
\affiliation{$^1$Department of Physics, Ko\c{c} University, \.{I}stanbul, Sar\i yer 34450, Turkey\\
$^2$College of Engineering and Natural Sciences, Bah\c{c}e\c{s}ehir University, Be\c{s}ikta\c{s}, Istanbul 34353, Turkey\\
$^3$Istituto Nazionale di Fisica Nucleare, Sezione di Milano, via Celoria 16, 20133 Milan, Italy\\
$^4$Dipartimento di Fisica ``Aldo Pontremoli", Universit{\`a} degli Studi di Milano, via Celoria 16, 20133 Milan, Italy\\
$^5$Centre for Theoretical Atomic, Molecular and Optical Physics, School of Mathematics and Physics, Queen's University Belfast, Belfast BT7 1NN, United Kingdom}

\begin{abstract}
We examine a simple scheme to generate genuine multipartite entangled states across disjoint qubit registers. We employ a shuttle qubit that is sequentially coupled, in an energy preserving manner, to the constituents within each register through rounds of interactions. We establish that stable $W$-type entanglement can be generated among all qubits within the registers. Furthermore, we find that the entanglement is sensitive to how the shuttle is treated, showing that a significantly larger degree is achieved by performing projective measurements on it. Finally, we assess the resilience of this entanglement generation protocol to several types of noise and imperfections, showing that it is remarkably robust.
\end{abstract}
\date{\today}
\maketitle

\section{Introduction}
It is well established that entangled systems allow for quantum enhanced information processing protocols. The advantage provided even from using two-qubit entanglement continues to drive studies into its generation, quantification, and characterization~\cite{HorodeckiRMP, CampbellPRA2009}. As the ability to control larger ensembles of quantum systems progresses, we are increasingly faced with the need to extend this framework to multipartite systems.

Genuine multipartite entanglement (GME) is an inherently more complex, but richer, phenomenon. To date, advances have been made in the understanding of three~\cite{DurPRA, SabinEPJD, GennaroPRA} and four qubit states~\cite{VerstraetePRA, GuhnePRARap}, while the characterization of GME in arbitrary systems is still an open question~\cite{MiyakePRA, deVincentePRA, JungnitschPRL, NovoPRA, SchwemmerPRL, LeviPRL, HuberPRA, EltschkaJPA}. Indeed, beyond information processing protocols, such tri- and quadripartite GME has proved to be a useful tool in condensed matter systems for exploring criticality~\cite{BayatPRL, GiampaoloPRA, HofmannPRB, CampbellPRA}. While entanglement in larger systems is comparatively less understood, it is nevertheless well established that such systems exhibiting GME can be useful, for example, $N$-qubit cluster states which allow for universal quantum computation~\cite{,ClarkNJP} and one-way quantum computation~\cite{raussendorf_one-way_2001}.

Multi-qubit $W$-states generated across qubit registers can be used for ensemble-based quantum memories~\cite{sangouard_quantum_2011}, quantum fuels~\cite{dag_multiatom_2016}, communication in quantum networks~\cite{LipinskaArXiv, JooNJP, ChiuriPRL, PuSciAdv} and distributed quantum computing, among other applications~\cite{gorbachev_multiparticle_2005}. Their scalability and robustness against local bit-flip noise, global dephasing, and particle loss make them advantageous resources as compared to other multipartite entangled states such as GHZ and cluster states. Although in terms of robustness, cluster states perform better than GHZ states, we lack scalable distinguishing analyzers for them, in both typical linear-optics based quantum communication set-ups and in atomic set-ups; while there are promising proposals for efficient analyzers of $W$-states for large networks \cite{ZhuSciRep}. Additionally, due to the inherent robustness of $W$-states and the non-local nature of their excitations, they are also understood to play a role in energy transport in light harvesting complexes~\cite{protein}. The scheme we propose for the generation of many-qubit $W$-states can be integrated into recent experimental quantum simulators of quantum biological systems~\cite{lightharvesting}.

An outstanding issue remains regarding the efficient generation and certification of GME. Several proposals have been developed in the literature each with their own associated strengths and drawbacks~\cite{ClarkNJP, GualdiNJP, SpillerPRA, YungPRA, YungQIC, BanchiPRL, CubittPRA,  HanOptEx2017, OzdemirNJP, Scheme1, Scheme2, Scheme3}, while there has been remarkable advances in the experimental generation of GME states~\cite{Exp1, Exp2, Exp3}. An additionally important issue relates to the ability to entangle disconnected parts of a quantum device which enhances their versatility~\cite{Buzek, HanggiArXiv, EmersonArXiv}. The inherent fragility of quantum systems requires that we develop methods to generate GME states involving minimal external control and initial resources. It is precisely in this direction that the present work progresses. We examine a simple protocol, which could be implementable with transmon qubits~\cite{transmon, WendinRPP2017, MezzacapoPRL2014, DalmontePRB2015, PaikPRL}, that uses a shuttle system which sequentially interacts with qubits inside disconnected registers. Projective measurements on this shuttle system are shown to significantly enhance the entanglement content. We show that an energy preserving interaction and suitable initial configuration is sufficient to ensure that a state with a large degree of GME is achievable. In addition we show this scheme readily generates states that are close to a perfect $W$-state, which is known to be a robust form of multipartite entanglement, while requiring significantly less control or resources compared to present techniques~\cite{ HanOptEx2017, OzdemirNJP,  Scheme1, Scheme2}. Finally, we examine the resilience of the protocol to several common sources of noise.

The generation of multi-qubit $W$-states is an active and challenging research field, due to the aforementioned vast range of applications that these states have in quantum information processing~\cite{raussendorf_one-way_2001, ClarkNJP, sangouard_quantum_2011, dag_multiatom_2016, LipinskaArXiv, JooNJP, ChiuriPRL, PuSciAdv, gorbachev_multiparticle_2005, ZhuSciRep, protein, lightharvesting, GualdiNJP, SpillerPRA, YungPRA, YungQIC, BanchiPRL, CubittPRA,  HanOptEx2017, OzdemirNJP, Scheme1, Scheme2, Scheme3, Exp1, Exp2, Exp3, Buzek, HanggiArXiv, EmersonArXiv}. One of the promising experimental results obtained to-date is the eight-ion $W$-state produced in an ion-trap quantum processor~\cite{haffner_scalable_2005}. The scheme is based upon sharing a common motional quantum between the ions with partial swap operations. The scheme is technically limited to scaling up to larger systems due to incomplete optical pumping. In our scheme we assume a common shuttle qubit between the two separate registers with swap operations. There are no external pumps and the shuttle is a separate physical entity instead of being another degree of freedom of the registers. The scenario here should produce a single excitation $W$-like state, though it is necessary for us to take into account possible errors and decoherence effects carefully. Our approach potentially offers significant technical advantages towards scalability and implementations, as we argue in the context of superconducting transmon qubits. It can be intuitively expected that the scheme should have further advantages in being able to fuse $W$-states with a higher number of excitations using more than one shuttle qubit with excitation conserving swap operations. We focus our attention here to the case of single shuttle and leave the cases of generation of other $W$-type (or Dicke) states, as well as GHZ type states to future studies; as they require addressing decoherence and errors in realistic settings separately.

\section{Set-up and Figures of Merit}
We consider a system composed of two disjoint registers of qubits labelled $r$ and $s$, respectively. We allow for qubits within a given register to interact with their nearest neighbors, however the two registers never directly communicate. Instead their mutual interaction is mediated by an ancillary shuttle qubit, $A$, which interacts with only one register qubit at a time. In what follows we assume all interactions are energy preserving such that the evolution operator can be conveniently expressed as a partial swap operation
\begin{equation}
\label{evolution}
U = \cos(\gamma) \openone +  i\sin(\gamma) \text{SWAP}.
\end{equation}
with $\text{SWAP}\!=\!\ket{00}\!\bra{00}+\ket{01}\!\bra{10}+\ket{10}\!\bra{01}+\ket{11}\!\bra{11}$ and $\gamma$ dictates the length of time, or equivalently the strength, of the interactions. As $A$ plays the role of a shuttle facilitating communication between $r$ and $s$, we will require that its interactions are short ranged and weak, while we will allow the interactions between qubits within a given register to take arbitrary values. In particular we will fix $\gamma\!=\!0.05$ for the shuttle-register interactions, while we will focus on the limiting cases of vanishing ($\gamma\!=\!0$) and strong ($\gamma\!=\!0.95\pi/2$) intra-register interactions. The choice of $\gamma\!=\!0.95\pi/2$ in the strong interaction case is to ensure the interaction results in an almost perfect SWAP operation within the registers. 

\begin{figure}[t]
\includegraphics[width=0.95\columnwidth]{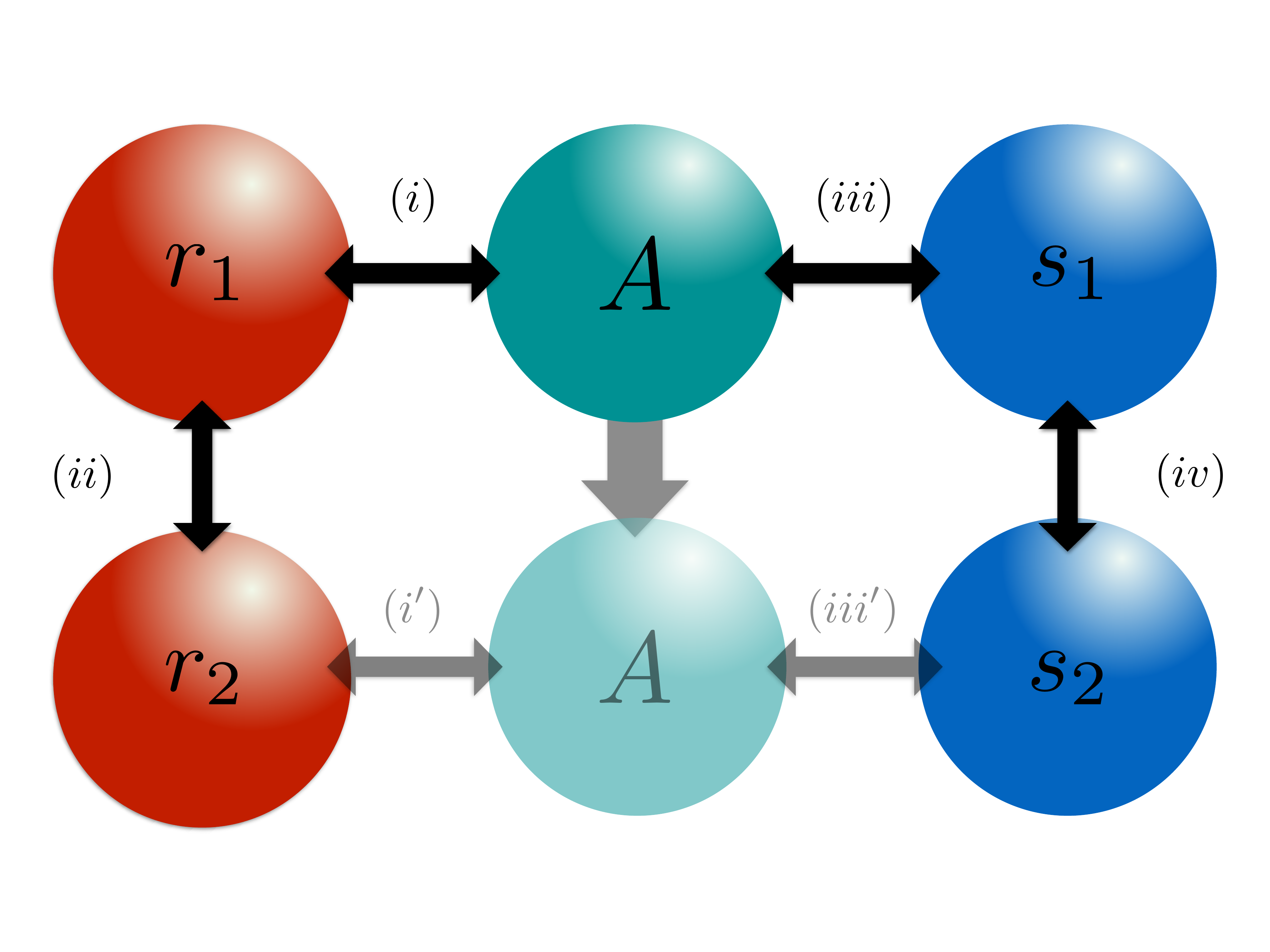}
\caption{Schematic of the setting. We consider two distinct registers of qubits, $r_1,~r_2$ and $s_1~,s_2$, whose mutual interaction is mediated by an ancillary qubit, $A$. As $A$ plays the role of a shuttle connecting disjoint parts of the system, we restrict its interactions to be weak, taking $\gamma\!=\!0.05$ in Eq.~\eqref{evolution}, while the interactions within a given register can be of arbitrary strength but are restricted to be between nearest neighbors. We discretize the time such that steps $(i)$-$(iv)$ represent the series of interactions realizing a single iteration of the process. The system exhibits quasi-periodicity after $\sim45$ iterations.} 
\label{fig1}
\end{figure}

In Fig.~\ref{fig1} we show a diagram of the considered scenario, where we have restricted the registers to be two qubits in length. In what follows we will exhaustively assess the entanglement properties in this particular setting, however we stress our results extend to larger register sizes. We will assume a discretized interaction time such that one step of the evolution corresponds to (up to) four interactions as labelled in Fig.~\ref{fig1} and, due to the shuttle-register interaction strength, the system exhibits a quasi-periodicity after $\sim\!\!45$ steps.

It should be noted that the scheme considered here shares important structural features with recently considered collision models (see e.g.~Refs.~\cite{FrancescoQMQM} and~\cite{ziman_description_2005} and references therein), which have been 
introduced to provide a microscopic framework for modeling open quantum systems. Note however that the perspective adopted in this paper is actually complementary to commonly considered collision models~\cite{FrancescoQMQM,ziman_description_2005}. 
In the context of our analysis, the shuttle corresponds to the open system, while one (or both) registers would represent an environment (here considered of finite and small dimension). While one is usually interested in the reduced system dynamics, here we look for the entanglement structure emerging within the environment in these microscopic models as a consequence of the repeated interaction with the system. 

Our collision model scheme can be compared to a variation of a quantum Turing machine~\cite{deutsch_quantum_1985}, where the shuttle qubit acts as a writing/reading head, moving across an information tape whose role is played by the register qubits. As a benchmark, let us consider an implementation based upon stochastic local operations assisted by classical communications (SLOCC) where operations can be partially successful~\cite{DurPRA}. We label the controllers of the head and the tape qubits as Alice and Bobs (Bob-1, Bob-2, ....). They are allowed to make local operations on their respective qubits and to communicate with the other controllers classically. When the head is on cell-$i$ then Alice and Bob-$i$ exchange the knowledge of the states of their qubits, tell the other Bobs to remain idle, then apply rotation on their qubits to exchange their states. If the rotation operations are not synchronised and suffer from time delays then swapping the states will be partially successful. Such a SLOCC implementation  will map product states into product states and cannot increase or produce  entanglement per se. 

Our collision model however is based on a quantum swap operation which is strictly non-local. Though it is not universal, we can use a typical physical model of the swap operation, often used in collision models, given by the Heisenberg exchange interaction
\begin{eqnarray}\label{eq:qSWAP}
\text{SWAP}=\exp{(iJ\tau \sigma_A\cdot \sigma_i/\hbar)},
\end{eqnarray}
where $J$ is the exchange constant, $\tau$ is the interaction time, $\sigma_A,\sigma_i$ are the Pauli spin operators for the shuttle (head) and the register (tape) qubits. Note that Eq.~\eqref{eq:qSWAP} agrees with the definition of a SWAP presented below Eq.~\eqref{evolution} at a suitable time up to a global phase factor. The free part of the Hamiltonian commutes with the exchange interaction as we assume identical qubits. Accordingly operations in the collision model scheme cannot raise the energy. Using Eq.~(\ref{eq:qSWAP}) in Eq.~(\ref{evolution}) and considering $n$ rounds of operations determined by $U^{\otimes n}$ we see that the terms in the evolution operator $U^{\otimes n}$ will be of the form
\begin{eqnarray}
\exp{(iH_{\text{eff}}\tau/\hbar)}
\end{eqnarray}
where 
\begin{eqnarray}
H_{\text{eff}}^{(m)}=J^m\sum_{i=1}^m\sigma_A\cdot\sigma_i,
\end{eqnarray}
with $m<n$. $H_{\text{eff}}$ is known as the Gaudin spin-star model~\cite{gaudin_diagonalisation_1976}. The Heisenberg exchange interactions between the register qubits and possible free evolutions of the non-colliding qubits where dephasing may occur during the intermediate stages are not included here. They are not part of the essential mechanism leading to $W$-type states in the scheme. Their positive and negative effects will be considered in the subsequent sections by numerical simulations.

According to the underlying central spin model physics, we predict that if we start with an excited shuttle qubit and register qubits in their ground states then the excitation will be distributed over all the sites and give the so-called $N$-qubit $W$-type entangled state. We will also show below that genuine symmetric $N$-qubit $W$-states can also be produced by further manipulation of the shuttle qubit. The central spin model can also be used to predict the existence of collapse and revival dynamics in the system as well as quasi-periodicity. We can write the effective Hamiltonians in terms of collective operators $a_q\sim \sum_i\sigma_i^-$ for the register qubits such that
\begin{eqnarray}
H_{\text{eff}}^{(m)}=J^m(\sigma_A^+a_q+\sigma_A^-a_q^\dag),
\end{eqnarray}
where we have dropped the term with $\sigma_A^z$. The evolution would then be the superposition of various excitation exchanges (Rabi oscillations) at
different discrete Rabi frequencies. Accordingly, we expect a collapse and revival dynamics for the system.

We will be interested in characterizing the entanglement properties within and across the two registers. The degrees of freedom associated to and any correlations shared with the shuttle will be neglected, which we will account for in two different processes. Firstly, we will simply trace out $A$'s degrees of freedom,
\begin{equation}
\label{traceRho}
\rho_T \equiv \rho_T^{r, s} =\text{Tr}_A \left[ \rho \right].
\end{equation}
Secondly, we will perform a projective measurement on the shuttle according to
\begin{equation}
\label{projectionRho}
\rho_P\equiv \rho_P^{r, s}  = \text{Tr}_A \left[ \frac{ \ket{0}_A\!\bra{0} \rho \ket{0}_A\!\bra{0} }{\text{Tr}\left[ \ket{0}_A\!\bra{0} \rho \ket{0}_A\!\bra{0} \right]} \right].
\end{equation}
Of course both procedures leave us with the reduced state of the two initially disconnected registers, however as we will show choosing either of them to treat the shuttle can have a significantly different effect on the entanglement content 
of the registers.

To quantify the entanglement we will rely on measures based on the positive partial transpose (PPT) criterion~\cite{HorodeckiRMP}, which faithfully detects entanglement in bipartite $2\times2$ and $2\times3$ dimensional systems. For two-qubit states the (unnormalized) negativity is defined as
\begin{equation}
\label{bipartneg}
{\cal N}_2=-\text{max} [0,\lambda_{\text{neg}}],
\end{equation}
where $\lambda_{\text{neg}}$ is the negative eigenvalue of the partially transposed density matrix, (for systems of more than two-qubits one would take the sum of the negative eigenvalues). While in general the PPT criterion is only a sufficient condition for entanglement, i.e. there may be PPT states which are entangled, it nevertheless can be extended to larger systems.

To quantify the GME we will use the generalization introduced in Ref.~\cite{JungnitschPRL}. By definition, if an arbitrary multipartite state can be written as a mixture of different possible bipartitions, it is called bi-separable. A GME state is defined as a state that does not admit such a bi-separable decomposition. In Ref.~\cite{JungnitschPRL} the problem of detecting GME is addressed by finding an entanglement witness which has a positive expectation value if a given density matrix can be written as a mixture of PPT states and has a negative expectation value if this is not possible. Note that the set of bi-separable states is a subset of PPT mixtures and therefore no GME state can be a PPT mixture. The key advantage of this approach is that constructing a witness with the aforementioned properties can be achieved by semidefinite programming~\cite{SDP}. Furthermore, it is then possible to define a GME measure by looking at the negative expectation value of the witness as it satisfies all the criteria to be an entanglement monotone. In fact, applied to two-qubit states it returns precisely the negativity defined in Eq.~\eqref{bipartneg}. Throughout this work we will adopt this approach in order to quantify the entanglement content among the registers. Computation of the measure is done using the code provided by the authors of Ref.~\cite{JungnitschPRL} available from \cite{footnote1}, together with the parser YALMIP~\cite{footnote2} and the solver SDPT3~\cite{SDPT_1, SDPT_2}.

\section{Entanglement Generation Across Disconnected Registers}
\label{clean}
\begin{figure}[t]
{\bf (a)} \\
\includegraphics[width=0.8\columnwidth]{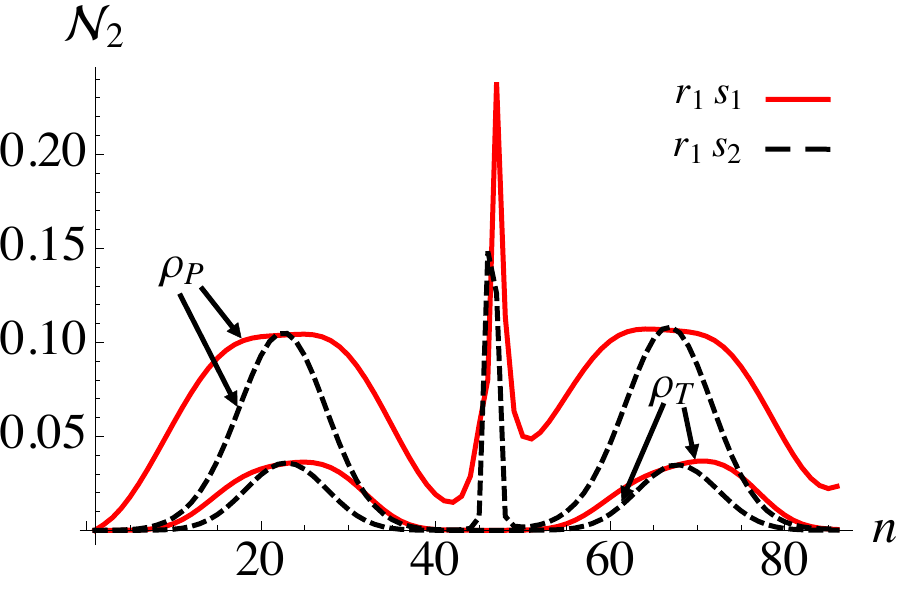}\\
{\bf (b)}\\
\includegraphics[width=0.8\columnwidth]{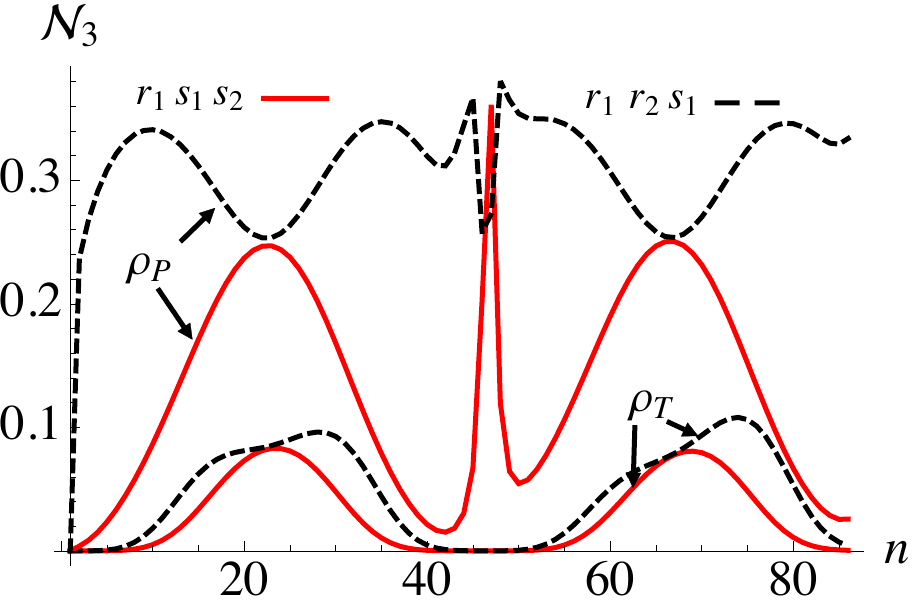}\\
{\bf (c)}\\
\includegraphics[width=0.8\columnwidth]{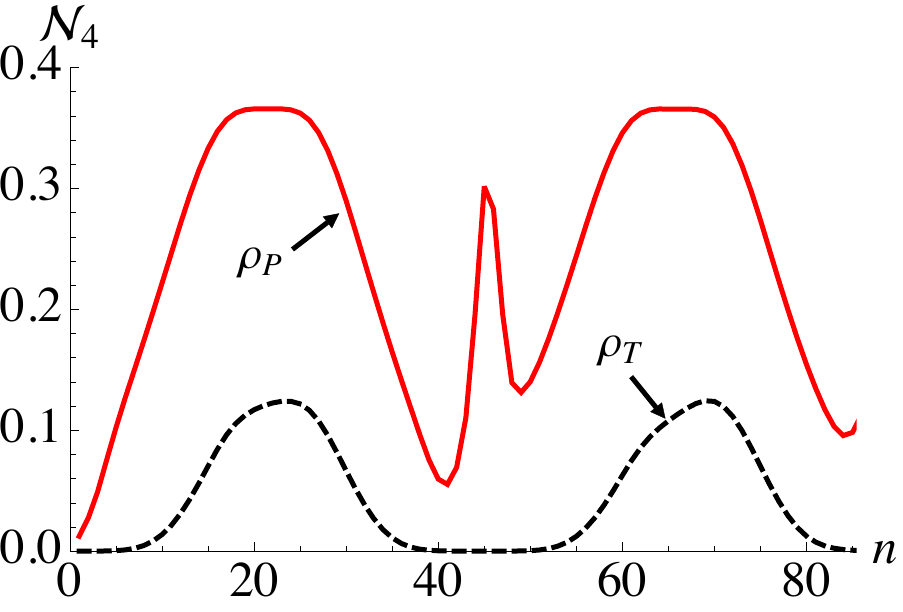}\\
\caption{{\bf (a)} Bipartite entanglement, {\bf (b)} Tripartite entanglement, and {\bf (c)} Quadripartite entanglement. In all panels register qubits are initialized in $\ket{0}$, while the shuttle's initial state is $\ket{1}$. The shuttle-register interaction is weak, $\gamma\!=\!0.05$ while we assume strong interactions within both registers, $\gamma\!=\!0.95\tfrac{\pi}{2}$. The upper curves in each panel correspond to a projective measurement performed on the shuttle, while the lower curves correspond to tracing over the shuttle's degrees of freedom.}
\label{fig2}
\end{figure}

We begin our analysis by considering the ideal situation wherein both registers are initially in the same factorized state, with each qubit in its respective ground state $\ket{0}$. In addition we fix the initial state of the shuttle to be $\ket{1}$ (some discussions regarding other initial states for the shuttle will be addressed in Sec.~\ref{Wstates}). In Fig.~\ref{fig2} we allow for strong interactions between the qubits within the two registers and assess the bi-, tri-, and quadripartite entanglement. As shown in panel {\bf (a)}, bipartite entanglement is generated between two qubits across the registers. We find the behavior is qualitatively the same regardless of whether we examine two qubits whose interaction is directly mediated, i.e. the shuttle interacts with the qubits sequentially as in the case of $r_1$ and $s_1$, or if there is a lag in the effective interaction between the spins, as in the case of qubits $r_1$ and $s_2$ and furthermore, the dynamical maxima achieved are comparable. We remark that a quantitative difference is present in the state of the registers depending on how the shuttle degrees of freedom are eliminated. A significantly larger amount of entanglement is achieved in the case of a projective measurement.

In Fig.~\ref{fig2} {\bf (b)} we turn our attention to GME present in tripartite reduced states. When the shuttle qubit's degrees of freedom are traced out we find that the tripartite GME behaves in a qualitatively similar manner to the bipartite entanglement. While there is some weak dependence on which three of the four qubits we analyze, nevertheless genuine tripartite entanglement is always established. Conversely, we find more significant differences when we perform the projective measurement on the shuttle. In this case the GME exhibited is much more sensitive to which of the three qubits we are analyzing. In the case of $r_1$, $s_1$, and $s_2$ we find a consistent behavior with all the previous cases, where the entanglement peaks once during the period of the evolution. Contrarily, for the tuple $r_1$, $r_2$, and $s_1$ the tripartite entanglement grows significantly faster in the initial stages and peaks twice during the same period. The order of interactions in the collision model scheme (cf.~Fig.~\ref{fig1}) leads to the ``lag'' in the rise of the entanglement among different partitions of the register qubits. Notice that these peaks occur when the complementary tri- and bipartite entanglements are relatively low. Additionally we find that in the middle of the period both tuples achieve comparable values for the GME in line with the behavior of the two-qubit states.

The quadripartite GME is shown in panel {\bf (c)}. We see that the overall qualitative behavior is consistent with that shown in panel {\bf (a)}. Again performing a projective measurement on the shuttle leads to a significant increase in the amount of GME generated across the registers. Interestingly, the quadripartite entanglement is quickly established in this case, while it requires almost an order of magnitude more steps if the shuttle is traced out.

\begin{figure}[t]
{\bf (a)} \\
\includegraphics[width=0.8\columnwidth]{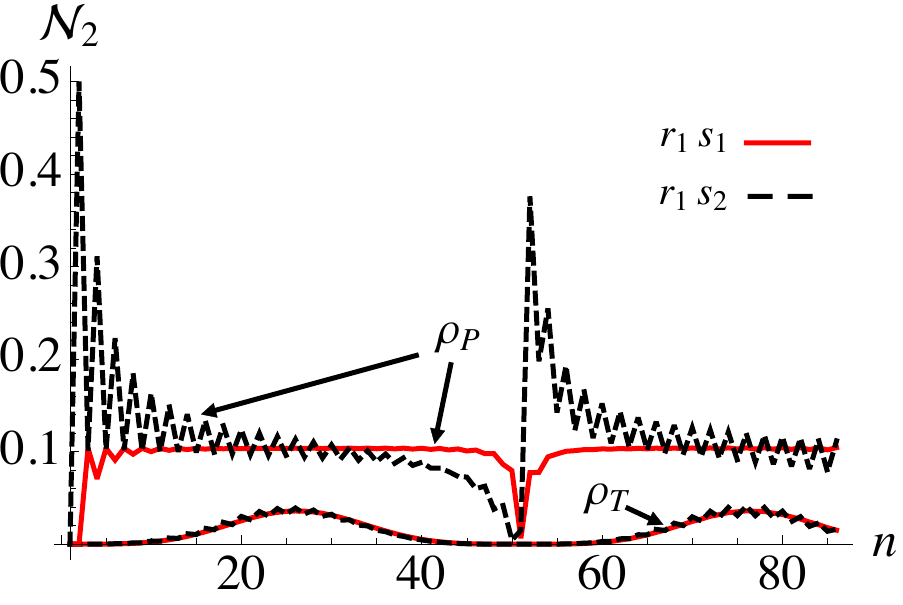}\\
{\bf (b)}\\
\includegraphics[width=0.8\columnwidth]{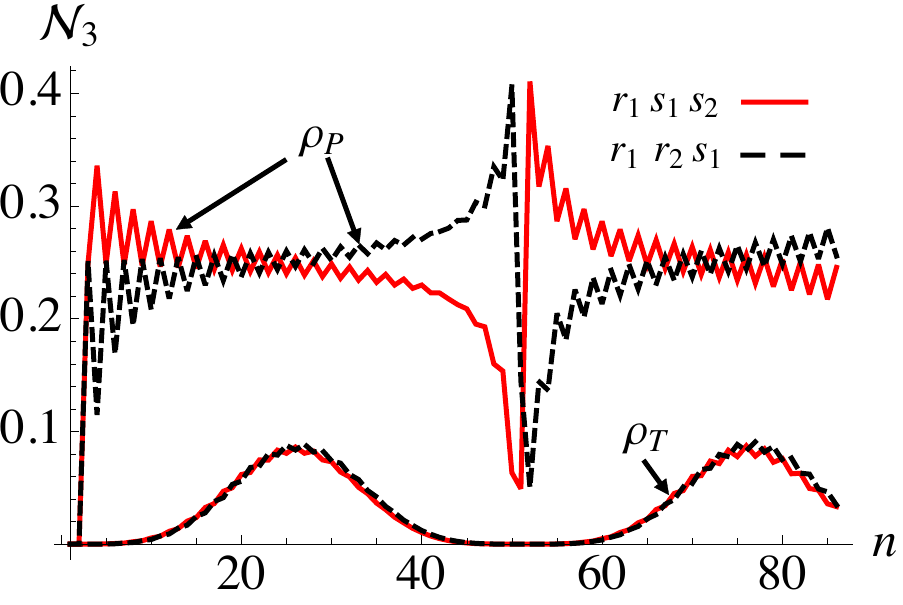}\\
{\bf (c)}\\
\includegraphics[width=0.8\columnwidth]{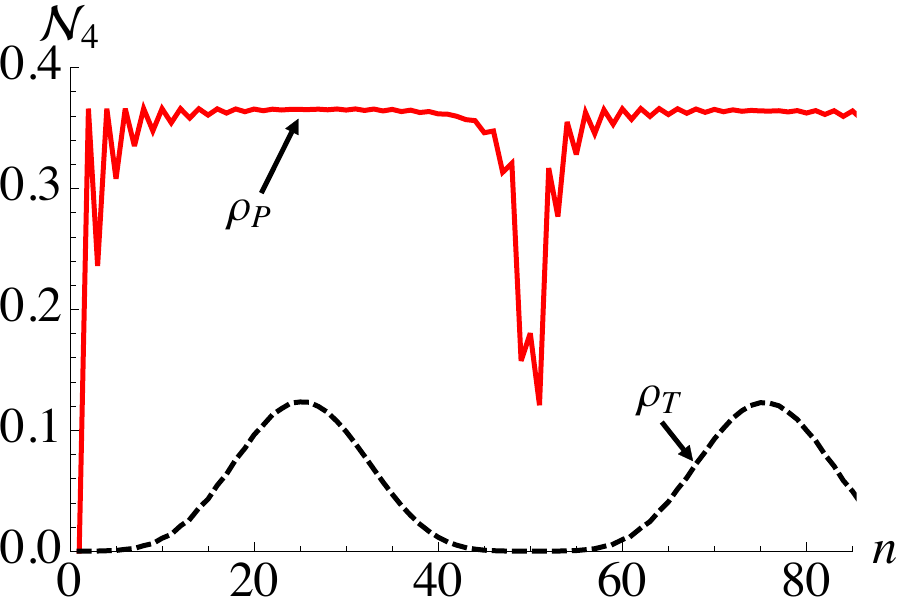}\\
\caption{As for Fig.~\ref{fig2}, except taking the interaction strength within the registers to be zero.}
\label{fig3}
\end{figure}

Fig.~\ref{fig3} examines precisely the same setting with one important difference: the qubits within a given register no longer directly interact with one-another. While it may be natural to assume that the strong interactions considered previously would facilitate the creation of entanglement, we find that this is not necessarily the case. When the shuttle is traced out we find that the overall features exhibited by the bi-, tri-, and quadripartite entanglements are largely unaffected, although evidently there are some minor quantitative and qualitative differences. The most remarkable difference is in the general behavior when a projective measurement on the shuttle qubit is performed. In panel {\bf (a)} we see the bipartite entanglement shared across the two registers between qubits $r_1$ and $s_1$ is almost constant, while the entanglement between $r_1$ and $s_2$ exhibits a more complex dynamics wherein it reaches much larger values initially followed by a jagged behavior, jumping significantly from one iteration to the next. In panel {\bf (b)} we see a similar jagged dynamics for the tripartite GME. It is interesting that once again there is a symmetry appearing between the two reduced tripartite states with the entanglement dynamics being almost perfectly antisymmetric. We find the quadripartite GME established is stable and behaves in a qualitatively identical manner to the bipartite entanglement shared between $r_1$ and $s_1$. After the first two iterations its value is already close to maximal, which is followed by a jagged transient before settling into an almost constant value for the rest of the period.

The main and non-trivial results here are that we can achieve the maximally entangled symmetric $W$-state in a shorter time, and freeze it, at least quasi-periodically. Fast generation and long time storage of many-qubit symmetric $W$-states are of broad interest for applications to quantum memories, quantum simulations, and quantum thermodynamics.

\subsection{Characterization of the entanglement structure}
\label{Wstates}
We can gain insight into these features by closer examination of the state generated during the dynamics. Due to the energy preserving nature of the interaction, the unitary dynamics of the overall system, and the fact that the only excitation in the total system is concentrated in the shuttle initially, it is easy to see that after at most two iterations the state of the shuttle+registers is
\begin{equation}
\begin{aligned}
\label{purestate}
\ket{\psi}_{r,A,s} =&~a_1 \ket{00001} + a_2 \ket{00010} + a_3 \ket{00100} \\ 
			   &+ a_4 \ket{01000} + a_5 \ket{10000},
\end{aligned}
\end{equation}
i.e. a five qubit `$W$-like' state, where we have used the ordering $\{ r_1$, $r_2$, $A$, $s_1$, $s_2 \}$ with $\sum_i \vert a_i \vert^2\!=\!1$ and $a_i\!>\!0$. While the coefficients are delicately dependent on the strength of the various interactions, the state is constrained to always take this form, which clearly exhibits GME. We can now clearly see the reason for projective measurements to lead to enhancements in the degree of entanglement established between the registers compared to tracing over $A$'s degrees of freedom. If we trace out the shuttle, the resulting state of the registers is 
\begin{equation}
\rho_T^{r,s} = \vert a_3 \vert^2 \ket{0000}\!\bra{0000} + (1-\vert a_3\vert^2) \ket{\psi_W^{r,s}}\!\bra{\psi_W^{r,s}},
\end{equation}
where 
\begin{equation}
\label{registerW}
\ket{\psi_W^{r,s}} =  b_1 \ket{0001} + b_2 \ket{0010} + b_3 \ket{0100} + b_4 \ket{1000},
\end{equation}
with $\sum_j|b_j|^2\!=\!1$. Conversely when a projective measurement on to the ground state of the shuttle is performed, the state of the registers is simply given by the pure GME state, Eq.~\eqref{registerW}. Thus, the entanglement content in both states is due to the same $W$-state structure present in Eq.~\eqref{registerW}. Tracing over $A$ simply leads to a classical mixture of an unentangled state with the four qubit entangled state of the register, and therefore must reduce the amount of GME present.

\begin{figure}[t]
{\bf (a)} \\
\includegraphics[width=0.8\columnwidth]{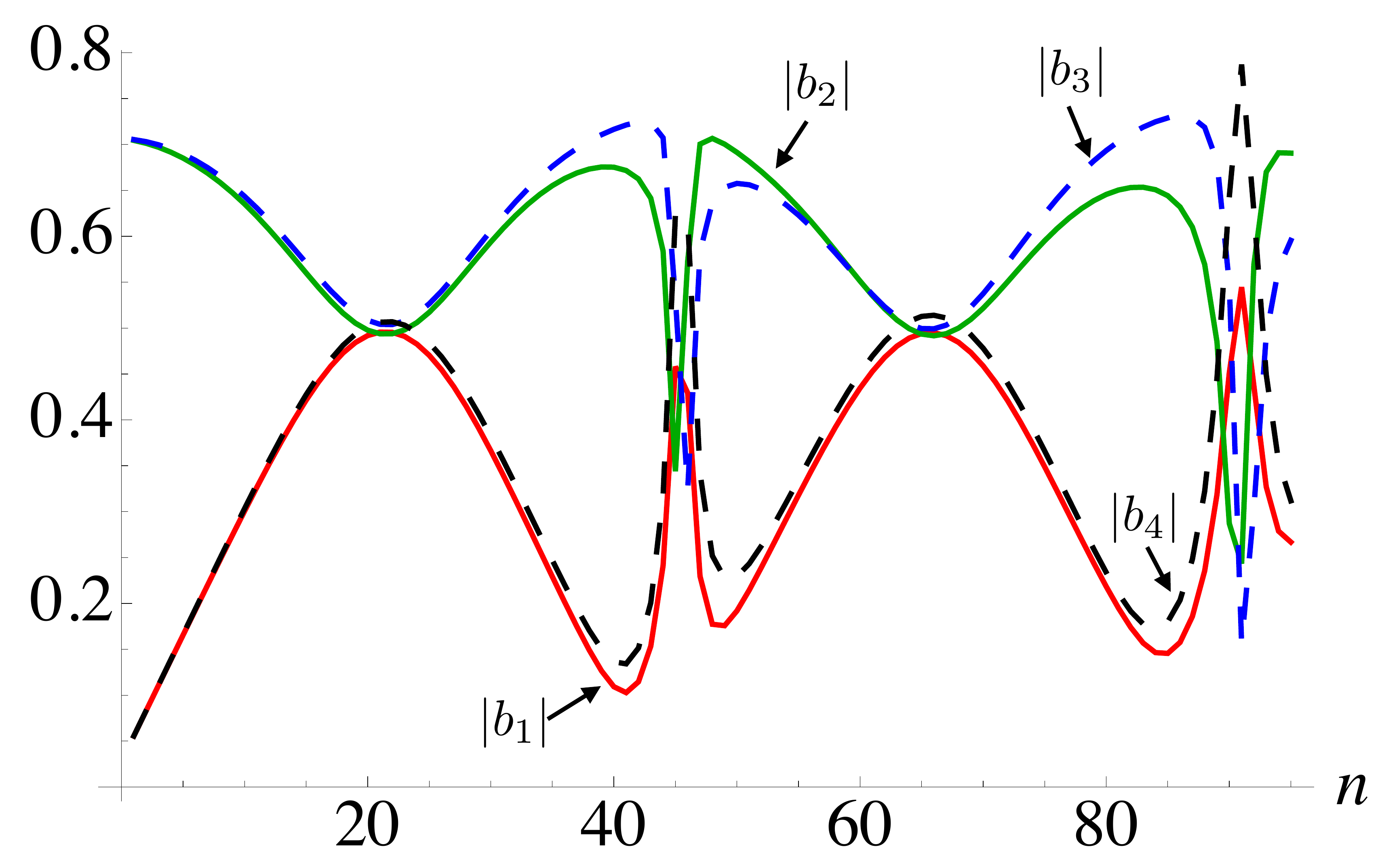}\\
{\bf (b)}\\
\includegraphics[width=0.8\columnwidth]{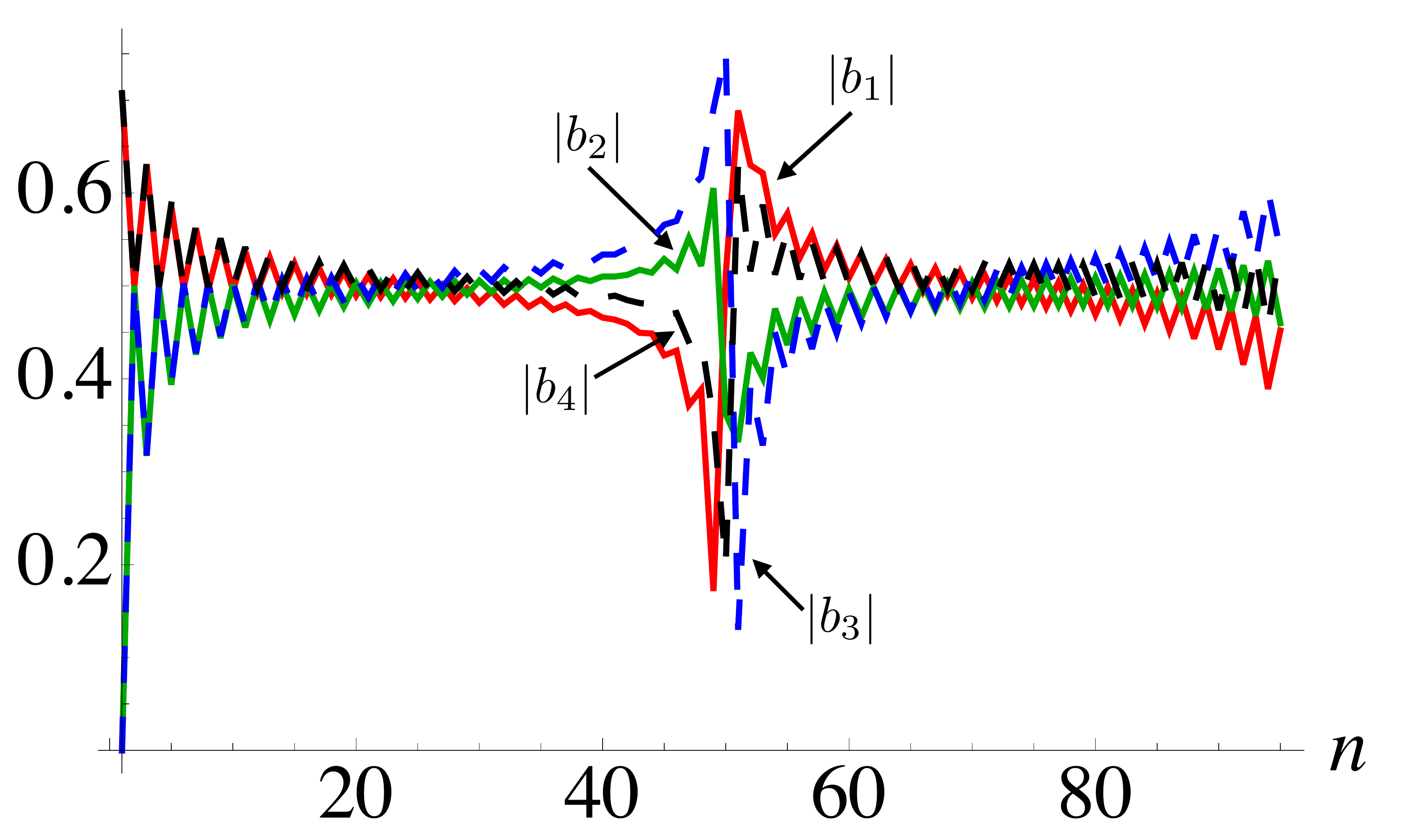}\\
\caption{Dynamics of the absolute values of coefficients $b_i$ in Eq.~\eqref{registerW} for {\bf (a)} strong and {\bf (b)} zero intra-register interactions. Note that for $b_i\sim0.5$ indicates that we have obtained a symmetric $W$-state.}
\label{fignew}
\end{figure}

In fact, closer examination of the coefficients $b_i$ in Eq.~\eqref{registerW} reveals that the states that achieve the maximal amount of GME in Figs.~\ref{fig2} {\bf (c)} and \ref{fig3} {\bf (c)} have a fidelity $>0.995$ with a perfect four-qubit $W$-state. This can also be seen by looking at the absolute value of $b_i$ throughout the dynamics, which is presented in Fig.~\ref{fignew}. The point where all coefficients become equal to $0.5$ is where the symmetric $W$-state is formed. It is possible to observe that for the strong intra-register interactions, Fig~\ref{fignew} {\bf (a)}, the register qubits quasi-periodically form a symmetric $W$- state for a short instance during the dynamics. On the other hand, for zero intra-register interactions, Fig~\ref{fignew} {\bf (b)}, the quasi-periodically formed $W$-state lasts for larger number of collisions, which can be viewed as freezing of the generated GME state. Considering the way in which the setup is initiated and the kind of interactions governing the dynamics, such a behavior may be expected to extend to arbitrarily sized registers. We have confirmed that similar fidelities occur between a $N$-qubit $W$-state and the dynamical state of our protocol for registers of up to five qubits in length (i.e. $N\!=\!10$), thus supporting the intuition that our scheme can efficiently generate GME states across disjoint systems with minimal requirements. Naturally, as the register size is increased the quasi-periodicity observed in our systems is affected, in particular for larger registers we observe in our numerical analysis that a slightly larger number of iterations of the process is required to come close to a $W$-state, with the overall periodicities of 45 iterations for two-qubit registers versus 56 for four-qubit registers. This indicates that the observed periodicity is related to both the geometry of the setting and the relative coupling strengths. Although one may argue that the scheme is expected to work this way given the initial state and the nature of the interaction, it is important to emphasise the necessity of the projective measurement on the shuttle in order to obtain the almost perfect symmetric $W$-state in the registers. A possible generalisation of the scheme may be to initialise the system with $k$-shuttles in the excited state instead of one and coherently distribute these excitations to obtain $k$-excitation Dicke state.

A similar analysis also elucidates the peculiar behavior exhibited near these periodic points in the dynamics when a projection is performed on the shuttle. It is easy to confirm that near these points the coefficient $a_3$ in Eq.~\eqref{purestate} is large, being $\gtrsim0.95$. The projective measurement therefore has a significant effect on the magnitude of the remaining coefficients such that small differences in their values are greatly enhanced in the resulting pure four-qubit state, Eq.~\eqref{registerW}. In Fig.~\ref{fig2} (Fig.~\ref{fig3}) when a spike (dip) appears in the GME, this corresponds to when the coefficients $a_1,~a_2,~a_4,~a_5$ are sufficiently small that from one iteration to the next the seemingly minor changes in amplitudes are drastically affected after the projective measurement on the shuttle. For the zero intra-register interactions case shown in Fig.~\ref{fig3}, the relatively flat behavior of the entanglement after a projective measurement is also succinctly explained in this way. If the registers do not directly exchange any information, then the excitation can only be moved throughout the system by the shuttle. As the shuttle interacts identically with every register qubit, each interaction is similar, although since the state of the shuttle changes after each interaction they are not equivalent. Regardless, the resulting coefficients of $a_1,~a_2,~a_4,~a_5$ remain largely comparable for each iteration, and therefore under the projective measurement we realise close to a perfect $W$-state in this case for almost the entire dynamics.

While the above analysis is restricted to the initial condition $\ket{00100}$, we find a qualitatively similar behavior for any pure initial state of the shuttle. If the shuttle is initially prepared in some superposition we find the only difference is a coherent contribution of $\ket{00000}$ in Eq.~\eqref{purestate}. It is easy to convince oneself that this additional contribution only reduces the observed amount of entanglement, while leaving the overall features largely intact. In particular the protocol still produces $W$-state GME across the registers, albeit reduced in magnitude.

\section{Robustness to Relevant Noise}\label{noise}
From the previous analysis it is clear that genuine multipartite entanglement can be created across disjoint quantum registers with only the use of a single ancillary shuttle qubit. While crucially the preceding section restricted to the idealized setting, in this section we examine the effect the most relevant sources of noise have on the generated entanglement. In particular we assess: {\it (i)} an imperfect implementation modelled via randomly missed shuttle-register interactions, {\it (ii)}  asymmetric thermal effects, and {\it (iii)} dephasing.

\subsection{Imperfect Implementation}
We begin assessing the robustness of the entanglement generation scheme by relaxing the requirement that all interactions faithfully occur. Here we will assume that the shuttle qubit may fail to interact with a register qubit with some probability, while the interactions within a given register take place without any failure.  This is justified for two reasons, firstly  it is reasonable to expect that the qubits in a register are close to one-another for example they could be spin ensembles with strong nearest neighbor coupling and therefore interaction effects may not be neglected. Secondly, as we have seen from the previous analysis the intra-register interactions play a significantly diminished role in the entanglement properties of the overall system. Therefore, in what follows we assume the same initial configuration as in the previous section, i.e. $\ket{00100}$, the same strong intra-register interactions, and that there is a probability, $p$, that there will be no interaction between the shuttle qubit and a register qubit at a given step, i.e. every shuttle-register interaction occurs with probability $(1-p)$. We perform a statistical average over many realisations for each value of $p$ to achieve good convergence.

\begin{figure}[t]
{\bf (a)} \hskip0.45\columnwidth {\bf (b)} \\
\includegraphics[width=0.5\columnwidth]{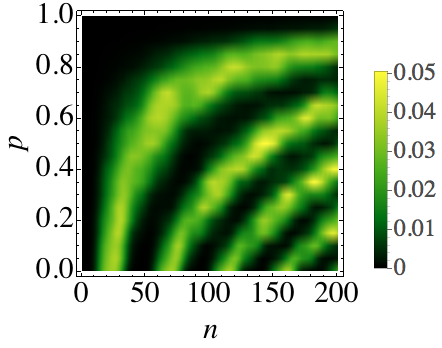}~\includegraphics[width=0.5\columnwidth]{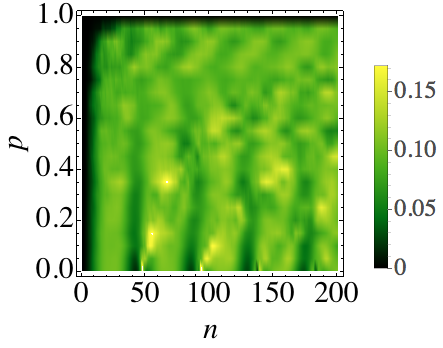}\\
{\bf (c)} \hskip0.45\columnwidth {\bf (d)} \\
\includegraphics[width=0.5\columnwidth]{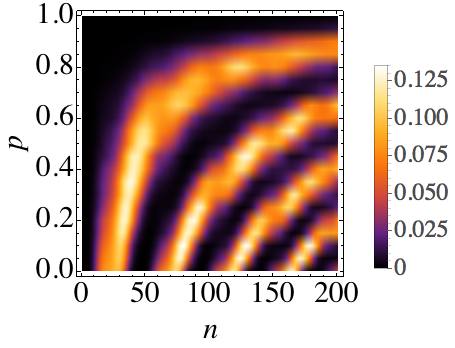}~\includegraphics[width=0.5\columnwidth]{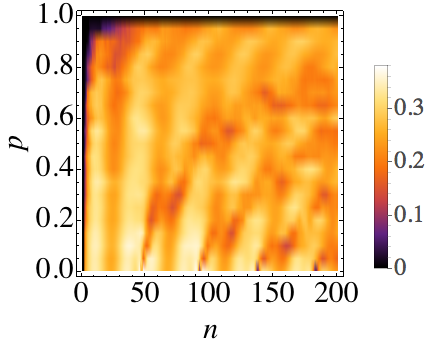}\\
{\bf (e)} \hskip0.45\columnwidth {\bf (f)} \\
\includegraphics[width=0.5\columnwidth]{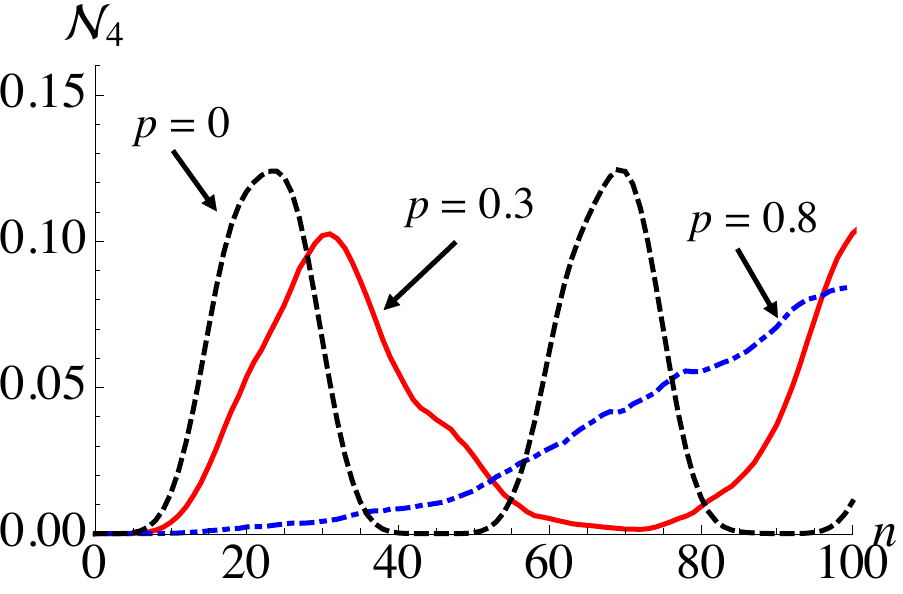}~\includegraphics[width=0.5\columnwidth]{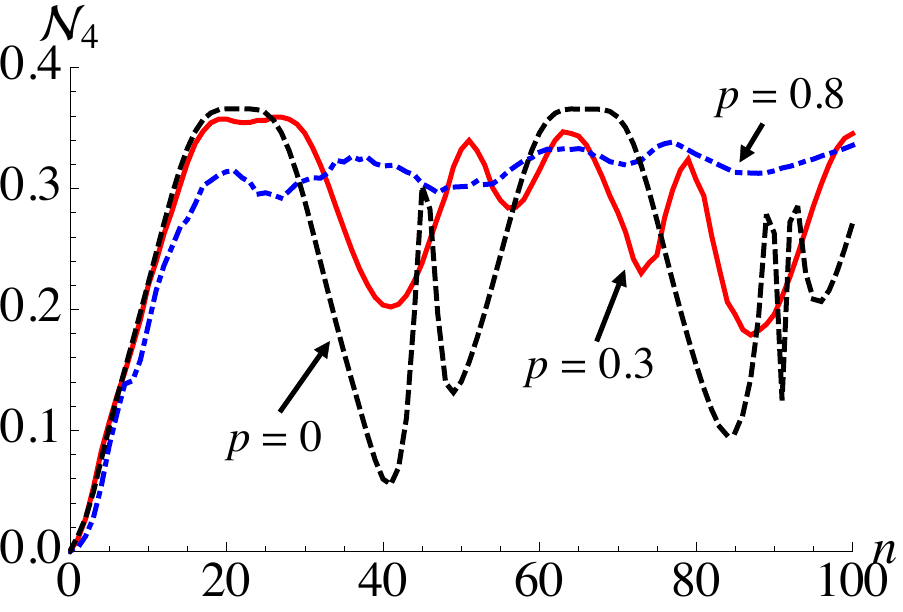}
\caption{Dynamical entanglement as a function of the probability, $p$, that the shuttle-register interaction is missed. {\bf (a)} Bipartite entanglement between $r_1$ and $s_1$ when the shuttle is traced out and {\bf (b)} after a projective measurement is performed on the shuttle. {\bf (c)} Tripartite negativity between qubits $r_1$, $r_2$, and $s_1$ when the shuttle is traced out and {\bf (d)} after a projective measurement is performed on the shuttle. {\bf (e)} Quadripartite entanglement after the shuttle is traced out for $p=0$, $p=0.3$, and  $p=0.8$ and {\bf (f)} after a projective measurement on the shuttle for $p=0$, $p=0.3$, and  $p=0.8$. In all panels we consider strong interactions within the registers, assume the register qubits' initial states to be $\ket{0}$, and the shuttle's initial state is $\ket{1}$.}
\label{fig4}
\end{figure}

In Fig.~\ref{fig4} we plot the bi-, tri-, and quadripartite entanglement generated as a function of $p$, all of which exhibit the same qualitative behavior. The left column shows the entanglement when the shuttle is traced out. For small $p$ the entanglement has the same quasi-periodicity shown in Fig.~\ref{fig2}. We see that while the rate of failure of interaction between the shuttle qubit and a register qubit, $p$, has strong affect on the quasi-periodic collision dynamics of the entanglement, it has neglgible effect on 
the amount of entanglement. In essence, a large probability to miss an interaction only delays the establishment of 
GME across the registers.

The right column corresponds to when the projective measurement on the shuttle is performed. In this case we find the entanglement is more robust to the failures in coupling the shuttle qubit to a register qubit interactions than in the case where the shuttle qubit is traced out. As evidenced in panels {\bf (b)}, {\bf (d)} and {\bf (f)} for moderate values of $p$ entanglement is still consistently created after only a few steps, and even when $p\!\sim\!0.9$ comparable amounts of entanglement are still generated after 30 iterations on average. Once again we find the entanglement generated after a projective measurement on the shuttle is performed is remarkably stable throughout the dynamics. Enhancement of the plateau over the number of collisions where the entanglement remains maximum suggests that increasing $p$ is a means for approximately turning off the shuttle-register interaction after the symmetric $W$-state is generated. This can be of practical significance as freezing the symmetric multi-qubit $W$-state would correspond to a long-life quantum memory for quantum information, thermodynamics, and simulation applications. While in Fig.~\ref{fig4} we have shown the results for the reduced states $r_1$-$s_1$ and $r_1$-$r_2$-$s_1$, we remark in line with Sec.~\ref{clean} the behavior is consistent for other suitable choices of qubits. 

\subsection{Thermal Effects}\label{temp}
We next examine how the initial temperature of the registers affects the ability to generate entanglement. In particular we will assume that the qubits in each register are initially in thermal equilibrium, i.e. $r_i$ ($s_i$) is described by the Gibbs state, $\rho_\text{th}=e^{-\sigma_z/T}/\mathcal{Z}$, at a temperature $T_1$ ($T_2$). In light of the previous analysis, we will again assume that all interactions take place faithfully, and we shall focus on the effect that this initial temperature has on the entanglement after 25 iterations, which corresponds to large amounts of entanglement in shared across all qubits, cfr. Sec~\ref{clean}.

\begin{figure}[t]
{\bf (a)} \\
\includegraphics[width=0.8\columnwidth]{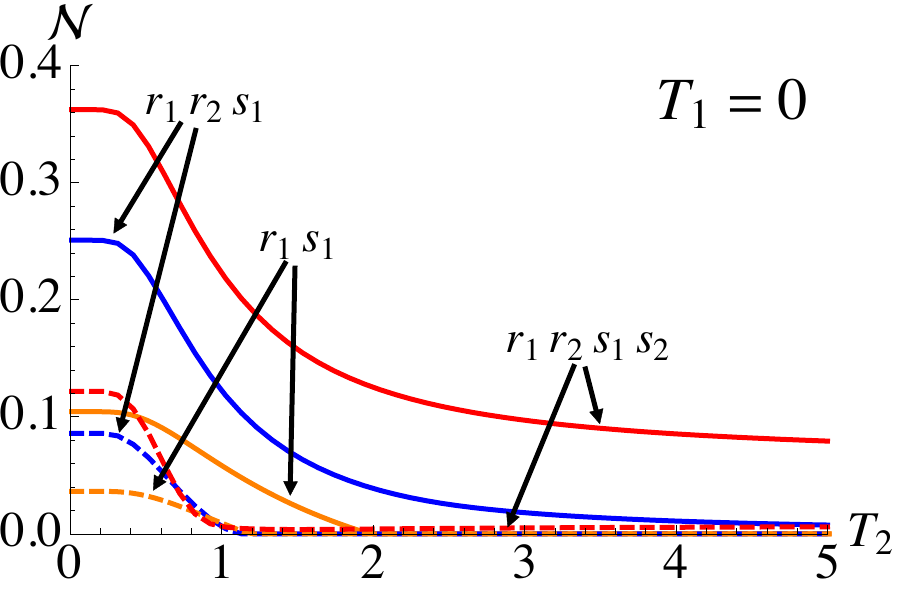}\\
{\bf (b)}\\
\includegraphics[width=0.8\columnwidth]{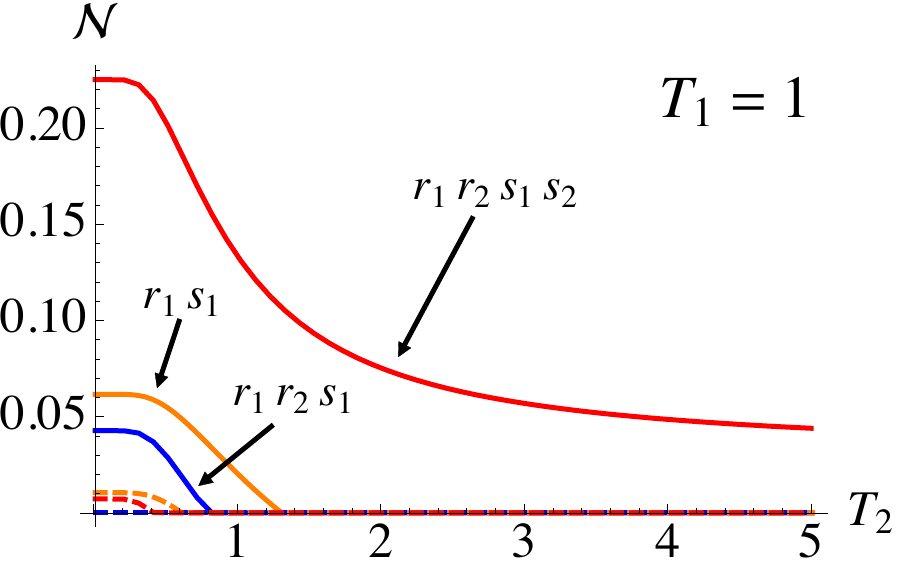}\\
\caption{Thermal effects on generated entanglement after $n=25$ iterations. Qubits within both registers are initially in Gibbs states at temperatures $T_1$ and $T_2$ for $r$ and $s$, respectively, while the shuttle is initially in its excited state $\ket{1}$. We consider strong interactions within the registers, $\gamma=0.95\tfrac{\pi}{2}$ and weak shuttle-register interactions, $\gamma\!=\!0.05$. In both panels solid curves correspond to a projective measurement on the shuttle, while dashed curves correspond to tracing over the shuttle's degrees of freedom. Temperature of $r$-register {\bf (a)} $T_1\!=\!0$ and {\bf (b)} $T_1\!=\!1$.}
\label{fig5}
\end{figure}

In Fig.~\ref{fig5} {\bf (a)} we consider the case when only register $s$ is thermal, while the qubits within $r$ are well isolated and therefore initialized in $\ket{0}$. Again we confirm that performing a projective measurement on the shuttle greatly enhances the observed entanglement. Indeed, for the bipartite entanglement established between $r_1$ and $s_1$ we see that it is significantly more resilient to the thermal effects than when $A$ is traced out, almost doubling the range of allowed temperatures for $T_2$ before the entanglement vanishes. This behavior is even more striking for the multipartite entanglement. For tripartite states, projective measurements on the shuttle lead to more than a five-fold increase in the resilience to the thermal effects, and this range is increased further for the quadripartite state. Indeed it is quite a remarkable feature that the GME established is consistently more robust to thermal noise than the entanglement of the reduced states, requiring higher temperatures before vanishing. In panel {\bf (b)} we see that if both registers are initially thermal, the overall effect is to reduce the achievable entanglement. In particular, when the shuttle is traced out all entanglement vanishes for small values of $T_2$. Conversely, projective measurements on the shuttle ensure a better resilience to the increased temperatures and the quadripartite entanglement remains non-zero for a wide temperature range.

A final interesting feature in Fig.~\ref{fig5} is the invariance of the observed entanglement for low temperatures. We clearly see that despite the mixing induced by the thermal noise, the GME established remains constant up to a given value of temperature, only after which this mixing becomes detrimental and leads to a decay in the entanglement. 

\subsection{Qubit Dephasing}
Finally, we consider a setting where independent dephasing channels act on the register qubits after they have interacted with the shuttle. Similar to the previous subsections, our aim is to determine how robust the generated entanglement is when the register qubits are subject to quantum noise. The Kraus operators describing the dephasing channel are 
\begin{align}
K_1 &= \sqrt{q} \begin{pmatrix}
    1 & 0 \\
    0 & 1
\end{pmatrix}
&
K_2 &= \sqrt{1-q} \begin{pmatrix}
    1 & 0 \\
    0 & -1
\end{pmatrix}.
\end{align}
This map can be viewed as leaving the qubit intact with probability $q$ and changing its relative phase with probability $1-q$. In a way the parameter $q$ controls how strong the dephasing process is affecting the register qubits.  Clearly, setting $q=1$ recovers the clean (noiseless) dynamics presented in Sec.~\ref{clean}. 
\begin{figure}[t]
{\bf (a)} \\
\includegraphics[width=0.8\columnwidth]{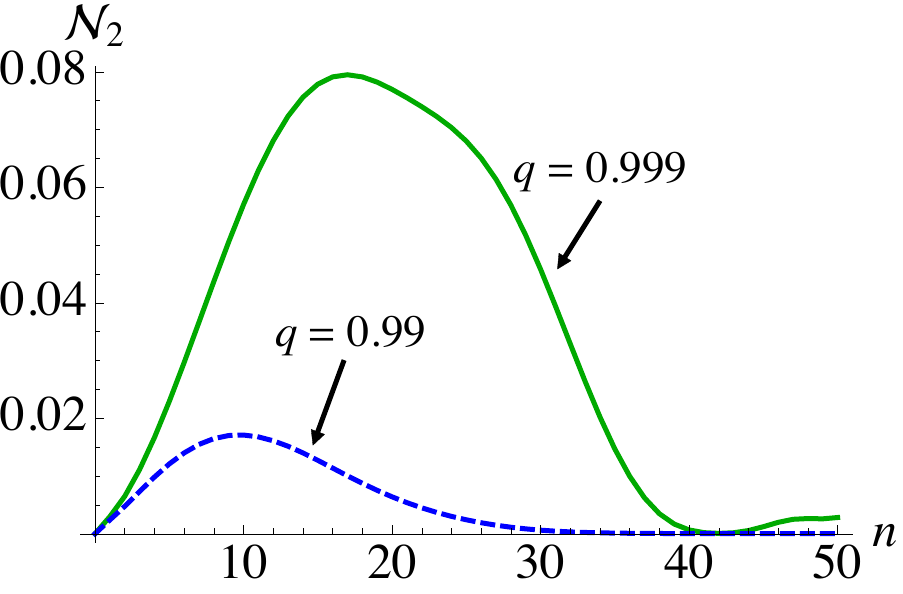}\\
{\bf (b)}\\
\includegraphics[width=0.8\columnwidth]{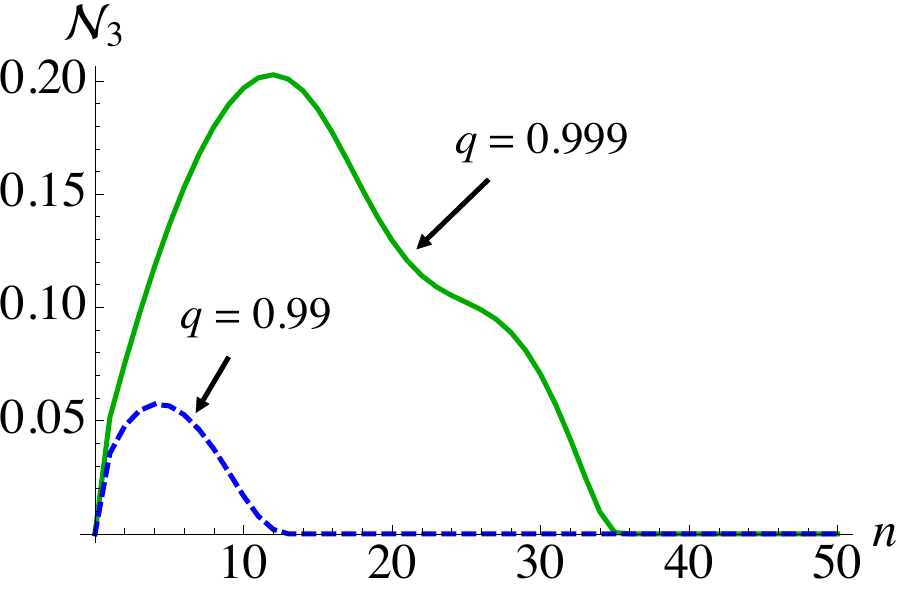}\\
{\bf (c)}\\
\includegraphics[width=0.8\columnwidth]{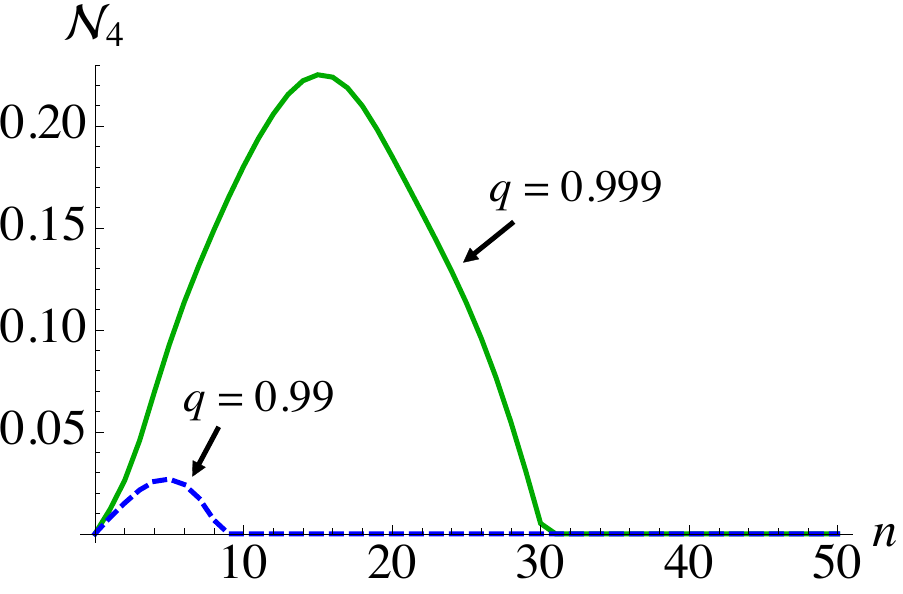}\\
\caption{Effects of qubit dephasing on {\bf (a)} Bipartite entanglement between $r_1$ and $s_1$. {\bf (b)} Tripartite entanglement between $r_1$, $r_2$ and $s_1$. {\bf (c)} Quadripartite entanglement across both registers. In all panels we assume strong interactions between register qubits $\gamma=0.95\tfrac{\pi}{2}$, weak shuttle-register interactions $\gamma=0.05$, all register qubits are initialized in state $\ket{0}$, shuttle qubit is initialized in state $\ket{1}$, and a projective measurement is performed on the shuttle.}
\label{fig6}
\end{figure}

In Fig.~\ref{fig6} we show the entanglement for different values of $q$ in the bi-, tri- and quadripartite cases for strong intra-register interactions and for a projective measurement on the shuttle qubit. Contrarily to the previous section, dephasing is shown to be detrimental to the entanglement, even a small amount of dephasing, i.e. $q\!\sim\!0.999$, is sufficient to significantly reduce the amount of entanglement present and all are null for $q\!<\!0.95$. It is worth noting that after a sufficient number of iterations all entanglement vanishes, first the quadripartite, followed by tripartite and finally the bipartite. Such as behavior is somewhat at variance with the other sources of noise previously considered where the system maintained a quasi-periodicity. 

\section{Possible Implementation}
A promising candidate system for the implementation of our protocol are transmon qubits due to their long coherence lifetimes and tuneable gate operations with short implementation times~\cite{transmon, WendinRPP2017, MezzacapoPRL2014, DalmontePRB2015, PaikPRL}. The state-of-the-art Josephson-junction transmon qubits have a coherence lifetime, $t_2$, of the order of $\sim\!10~\mu\text{s}$, with the typical interaction time between the qubits being $\sim\!10~\text{ns}$ ~\cite{PaikPRL}. The switching on and off of these interactions plays the role of our qubit collisions. Considering one iteration of our model consists of at least four collisions, the time needed to complete $50$ collisions is around $100~\text{ns}$, which is still two orders of magnitude shorter than the $t_2$ time. The dephasing values, $q$, corresponding to the time scales mentioned above can be found from the relation $q=(\exp(-t/2 t_2)+1)/2$, which for transmon qubits gives us $q_t\!\approx\!0.9975$. Clearly this value of $q_t$ lies between the $q$ values reported in Fig.~\ref{fig6} and hence implementing our scheme with transmon qubits, all bi-, tri- and quadripartite entanglement are present among the register qubits. Certainly, this conclusion is valid assuming that the register qubits are at zero temperature. However, the operating temperatures of transmon qubits with the considered specifications are around $10$ mK ~\cite{PaikPRL}, which is well within the temperature window where all types of entanglement are present, as shown in Sec~\ref{temp}.

One possible source of noise that have not been considered in Sec.~\ref{noise} is the imperfect initialization of the shuttle qubit in our scheme; we assume that our shuttle always starts in a pure state. While this assumption is evidently not always justified, for our envisaged implementation using transmon qubits realized in circuit QED setups, the initialization of all qubits, including the registers, to desired pure states can be performed by the fast reset of qubits to their ground state followed by application of suitable gates~\cite{BarendsNature}. The success of the initialization is therefore limited by the purification fidelity of the ground state preparation. If the reset is employed passively, such that the qubit waits to reach equilibrium with the ambient temperature of the experimental setup, then the time scale is determined by the relaxation time. If faster reset times are required, active refrigeration schemes can be employed~\cite{TanNatComm, ValenzuelaSci}. Early experimental implementations of active purification with autonomous feedback techniques reported high fidelities (more than 99.5\%) for the preparation of the ground state in short times (less than 3 s) \cite{GeerlingsPRL} and recently these results have been improved using 3D transmon qubits~\cite{JinPRL}.

Another possible platform to implement the present scheme is molecular nanomagnets, which are also known to have $t_2$ times at the order of $\sim\!10~\mu\text{s}$ \cite{mnm1}. It has been shown that it is possible to achieve a switchable coupling among them, enabling the possibility to implement two-qubit gates with application times around $\sim\!10~\text{ns}$ \cite{mnm2}. Furthermore, such systems can be combined to build scalable architectures, such as the setup considered in this work. Therefore, due to the similarity in the time scales with transmon qubits, the applicability of our protocol outlined previously also applies to molecular nanomagnets.

Current literature on the generation of $W$-states rely on different techniques. One of the well-known schemes is the cavity-fiber-cavity systems~\cite{Scheme1, Scheme2}, where the entangled states are generated by coupling the atoms or molecules inside a cavity through an optical fiber and attempt to engineer central spin (qubit-star or spin star) models~\cite{gaudin_diagonalisation_1976} indirectly by fine tuning the model parameters. These proposals naturally require a continuous physical coupling between the cavities, whereas we consider discrete interactions among the shuttle and register qubits with less control over the system parameters. Another well-established method is based on fusion techniques~\cite{ HanOptEx2017, OzdemirNJP}, where two or more smaller sized entangled states are sent through the fusion operation and, with a certain probability, create a larger sized $W$-state. Our scheme substantially differs from them, mainly due to the fact that it does not require any previously entangled resources. Instead, we initiate our protocol with a product state and the energy preserving character of the interaction between the shuttle and registers unitarily drives the system close to a $W$-state of the size of registers. Therefore, the model presented here is capable of generating $W$-states with limited control and resources compared to other techniques in the field. 

\section{Discussion and Conclusions}
We have assessed a simple entanglement generation scheme that creates robust genuine multipartite entanglement (GME) in disconnected systems. We have shown that through rounds of interactions with a shuttle system, which mediates effective coupling between disjoint registers, all the constituent qubits become entangled. In the case of `clean' registers, with all qubits initially in their respective ground states, it is sufficient to engineer an energy preserving interaction between a shuttle, initialized in its excited state, and the qubits in the register. The degree of entanglement, and most crucially the symmetric $W$-state generation, was shown to be sensitive to the way in which the shuttle is manipulated: we established that projective measurements lead to a significant increase in the GME compared to simply tracing out the shuttle's degrees of freedom, which can be understood from the convexity of entanglement, such that the produced state comes very close to the maximally entangled symmetric many-qubit $W$-state. We further showed that the entanglement generated was not significantly affected by additional interactions occurring within the registers. We extended our analysis to take into account several sources of noise in the process, specifically, a failure to couple the shuttle qubit to a register qubit during a given collision with some probability, initial thermal states of the register qubits, and dephasing in the register qubits. We showed that the proposed scheme is robust under these imperfections; the most degrading effects are due to dephasing, while increasing absence of shuttle qubit and register qubit interactions can be used to freeze the generated W state for quantum memory applications.

Recent work has shown that controlling the various interactions between the microscopic constituents in these collision 
models allows to explore in a robust and systematic way both Markovian, i.e. memoryless, and non-Markovian dynamics~\cite{BarisPRA, RuariPRA, StrunzPRA2016, LorenzoPRA2015, Vacchini2016b, Lorenzo2016a, Lorenzo2017a, JinNJP, CampbellPRA2018}. This suggests that an extension of our analysis could also be used to explore the possible relevance of non-Markovianity in entanglement generation.

\acknowledgements
MP, B\c{C} and \"{O}EM acknowledge support from Royal Society Newton Mobility Grant (Grant No. NI160057). B\c{C} and \"{O}EM acknowledge support from University Research Agreement between Ko\c{c} University and Lockheed Martin Chief Scientist's Office. BV acknowledges support from the EU Collaborative Project QuProCS (Grant Agreement No. 641277) and FFABR. \"{O}EM acknowledges support by TUBITAK (Grant No. 116F303) and the EU-COST Action (CA15220). MP acknowledges support from the DfE-SFI Investigator Programme (Grant No. 15/IA/2864) and the H2020 Collaborative Project TEQ (Grant Agreement No. 766900).

\bibliography{GME_collisional}

\end{document}